\documentclass[a4paper,reqno]{amsart}

\usepackage[utf8]{inputenc}
\usepackage[english]{babel} 

\usepackage{amsmath,amssymb,amsfonts,amsthm}
\usepackage{mathrsfs, esint, dsfont}
\usepackage{moreverb,rotating,graphics,float, subcaption, xcolor}
\usepackage[colorlinks, linkcolor={blue}, citecolor={red}]{hyperref}
\usepackage{framed,fancybox}
\usepackage{enumerate}
\usepackage{csquotes}
\usepackage{float}
\usepackage{slashed}

\newtheorem{theorem}{Theorem}

\newtheorem{remark}[theorem]{Remark}
\newtheorem{lemma}[theorem]{Lemma}

\newcommand\1{{\mathds 1}}

\def\N{{\mathbb N}}

\def\Q{{\mathbb Q}}
\def\R{{\mathbb R}}

\def\Z{{\mathbb Z}}

\def\be{{\mathbf e}}

\def\bk{{\mathbf k}}

\def\bt{{\mathbf t}}

\def\bx{{\mathbf x}}

\def\by{{\mathbf y}}

\def\rd{{\mathrm{d}}}
\def\re{{\mathrm{e}}}
\def\ri{{\mathrm{i}}}

\def\cB{{\mathcal B}}
\def\cC{{\mathcal C}}

\def\cN{{\mathcal N}}

\newcommand{\sC}{\mathscr{C}}

\newcommand{\eps}{\epsilon}

\newcommand{\Tr}{{\rm Tr}}
\newcommand{\VTr}{\underline{\rm Tr}}

\newcommand{\WS}{\mathbb{K}}
\newcommand{\BZ}{{\mathbb{K}^*}}
\newcommand{\BZy}{{\mathbb{K}^*_2}}

\newcommand{\bra}{\langle}
\newcommand{\ket}{\rangle}

\renewcommand{\epsilon}{\varepsilon}

\newcommand{\review}[1]{{#1}}

\def\bulk{{\rm bulk}}
\def\edge{{\rm edge}}

\author{David Gontier}
\address{CEREMADE, University of Paris-Dauphine, PSL University, 75016 Paris, France}
\email{gontier@ceremade.dauphine.fr}

\title[]{Spectral properties of periodic systems cut at an angle}

\date{\today}

\begin{document}

\begin{abstract}

We consider a semi-periodic two-dimensional Schrödinger operator which is cut at an angle. When the cut is commensurate with the periodic lattice, the spectrum of the operator has the band-gap Bloch structure. We prove that in the incommensurable case, there are no gaps: the gaps of the bulk operator are filled with edge spectrum.

\bigskip

\noindent \sl \copyright~2021 by the author. This paper may be reproduced, in its entirety, for non-commercial purposes.
\end{abstract}

\maketitle

\section{Introduction}

We study the spectral properties of half-periodic materials, when this one is cut along any line. Such materials are represented by Schrödinger operators of the form
\[
    H^D[\theta] := - \Delta + V_\theta, \quad \text{acting on $L^2(\R^2_+)$, with Dirichlet boundary conditions},
\]
where $\R^2_+  := \R_+ \times \R$ is the (right) half plane, and where $V_\theta$ is an $\theta$-rotated version of some $\Z^2$-periodic and bounded potential $V$, that is
\[
    V_\theta(x) := V \left( R_\theta^{-1} x  \right), \quad R_\theta := \begin{pmatrix}
        \cos \theta & -\sin \theta \\
        \sin \theta & \cos \theta
    \end{pmatrix}.
\]
When there is no cut, the bulk operator $H[\theta] := - \Delta + V_\theta$ is a rotated version of $H := H[0]$. Its spectrum $\sigma_\bulk$ is independent of $\theta$, and has a band-gap structure by Bloch theory.

When $\tan (\theta)$ is rational, of the form $\tan \theta = \frac{p}{q}$, the edge operator $H^D[\theta]$ is still periodic in the $x_2$ direction (with period $(p^2 + q^2)^{1/2}$). One can apply partial Bloch theory in this direction, and obtain that its spectrum has again the band-gap structure. This spectrum usually differs from $\sigma_\bulk$ due to the presence of edge modes. This is described by the edge spectrum
\[
    \sigma_\edge[\theta] := \sigma \left( H^D[\theta] \right) \setminus \sigma_\bulk.
\]
When $\tan (\theta)$ is not rational, one cannot apply Bloch theory. We prove the following.

\begin{theorem} \label{th:main}
    If $\tan \theta \notin \Q$, then there is $\Sigma \in \R$ such that $\sigma(H^D[\theta]) = [\Sigma, \infty)$.
\end{theorem}

In other words, in the incommensurable case, all gaps of $\sigma_{\bulk}$ are filled with edge spectrum. This extends the previous work by Hempel and Kohlmann~\cite{hempel2011spectral, hempel2012dislocation}, where the authors proved this filling gap phenomenon in the limit $\theta \to 0$. Here, we slightly modify their proof to handle all $\tan (\theta)$ irrational. The main tool that we use is the existence of a spectral flow when $\tan (\theta)$ is rational, and a limiting argument. Apart for the last part, we mostly follow the arguments by Hempel and Kohlmann in~\cite{hempel2011spectral, hempel2011variational, hempel2012dislocation, hempel2015bound}.

\review{
    When $\tan (\theta)$ is rational, the edge spectrum is absolutely continuous by Bloch theory (see also~\cite{davies1978scattering}). The corresponding eigenspace therefore describes modes that can propagate along the cut. However, when $\tan (\theta)$ is not rational, it is unclear what the nature of this edge spectrum is. Some part (or all of it) could be pure point, hence describing Anderson-like trapped modes which are localized near the cut. We do no investigate this interesting question in the present paper.
}

We choose for simplicity Dirichlet conditions at the cut $\{ 0 \} \times \R$, but the result can be generalized to other boundary conditions, such as Neumann boundary conditions. We only require that the condition in Eqn.~\eqref{eq:diff_res_compact} below holds.  We treat the case of domain walls in Section~\ref{ssec:domain_wall}.

\review{
    \begin{remark}
    In the special case of Dirichlet boundary conditions, the $\Sigma$ appearing in Theorem~\ref{th:main} is independent of $\theta$, and equals the infimum of the bulk spectrum $\Sigma = \inf \sigma_{\rm bulk}$. Indeed, adding Dirichlet boundary conditions corresponds to considering a smaller core domain for the forms, so $\inf \sigma(H^D) \ge \inf \sigma(H)$. The fact that we have equality is proved below. We thank the anonymous referee for this remark.
\end{remark}
}


\section{Background}
\subsection{Bulk Hamiltonian}

Let $V \in L^\infty(\R^2)$ be a bounded potential which is $L\Z^2$-periodic (at the end, we take $L = (p^2 + q^2)^{1/2}$ in the rational case $\tan \theta = \frac{p}{q}$), and let $H := - \Delta + V$ be the corresponding Hamiltonian. Since $H$ commutes with $L\Z^2$-translations, we can perform a Bloch decomposition~\cite[Chapter XIII]{reed1978analysis}, and write
\[
H = \int^\oplus_{\BZ} H_\bk \rd \bk, 
\]
where $\BZ := [- \frac{\pi}{L},  \frac{\pi}{L} ]^2$ is the Brillouin zone, and where $H_\bk := - \Delta +V$ is acting on $L^2(\WS)$, with $\WS := [0, L]^2$ the Wigner-Seitz cell. The operator $H_\bk$ has the $\bk$-dependent domain representing the usual $\bk$-quasi-periodic boundary conditions.

The map $\bk \mapsto H_\bk$ is $(2 \pi/L) \Z^2$-periodic. Each $H_\bk$ is compact resolvent, and we denote by $\varepsilon_{1 \bk} \le \varepsilon_{2 \bk } \le \cdots$ the eigenvalues of $H_\bk$, counting multiplicity. The maps $\bk \mapsto \varepsilon_{n \bk}$ are continuous and $(2 \pi/L) \Z^2$-periodic. This gives the usual band-gap structure of the bulk spectrum 
\[
    \sigma_{\rm bulk} = \sigma(H) = \bigcup_{\bk \in \BZ} \sigma(H_\bk) = \bigcup_{n = 1}^\infty \bigcup_{\bk \in \BZ}  \{ \varepsilon_{n \bk}\}.
\]

An energy $E$ is in a spectral gap of $H$ iff there is an integer $N$ so that
\[
\forall \bk \in \BZ, \quad \varepsilon_{N \bk} < E < \varepsilon_{N+1, \bk}.
\]
In what follows, we denote this integer $N$ by $\cN(E)$. It is the number of Bloch bands below the energy $E$. It is also the number of particles per unit cell for the state $\gamma_E := \1( H < E)$, in the sense that the trace per unit cell of $\gamma_E$ is
\[
     \VTr(\gamma_E) :=  \dfrac{1}{| \BZ |} \int_{\BZ} \Tr ( \gamma_\bk  ) \rd \bk = \cN(E),
\]
The number $\cN(E)$ is independent of $E$ in an open gap $g$ of $\R \setminus \sigma_\bulk$, and we sometime write $\cN(g)$ for $E \in g$.

\subsection{Dislocated Hamiltonians} We now focus on a {\em dislocated} version of the bulk operator. For $t \in \R$, we set
\[
    V_t(\bx) = V(\bx - t \be_1).
\]
By periodicity of $V$, the map $t \mapsto V_t$ is $L$-periodic. We also introduce the dislocated potential 
\begin{equation} \label{eq:def:Wt}
    W_t(\bx) := \left[ V(\bx) \1 (x_1< 0) + V_t(\bx) \1(x_1 > 0) \right],
\end{equation}
which represents a dislocation between the potential $V$ on the left side $x_1 < 0$, and a translated (dislocated) version of it $V_t$ on the right $x_1 > 0$. The {\em dislocated edge Hamiltonian} is defined by
\[
    H^\sharp(t) := - \Delta + W_t, \quad \text{acting on $L^2(\R^2)$, with domain $H^2(\R^2)$}.
\]
The spectral properties of such Hamiltonians have been studied {\em e.g.} by Davies and Simon~\cite{davies1978scattering} and Hempel and Kohlmann~\cite{hempel2011variational, hempel2011spectral, hempel2012dislocation, hempel2015bound}.
The map $t \mapsto H^\sharp(t)$ is $L$-periodic. When $t \in L \Z$, we recover the bulk Hamiltonian~$H$ (there is no dislocation), whose spectrum is $\sigma_{\rm bulk}$. However, when $t$ increases from $0$ to $L$, this spectrum may vary, as we explain now.

For all $t \in \R$, the operator $H^\sharp(t)$ is periodic in the $x_2$-direction, and we can write its partial Bloch expansion
\[
    H^\sharp(t) = \int^\oplus_{\BZy} H^\sharp_k(t) \rd k,
\]
where $\BZy := \left[\frac{-\pi}{L}, \frac{\pi}{L} \right]$ is the Brillouin zone in the $x_2$-direction only, and where $H^\sharp_k(t) = - \Delta + W_t$ acts on the tube $L^2\left( \R \times [0, L]  \right)$, and with the $k$-dependent domain representing the $k$-quasi-periodic boundary condition in the $x_2$-direction. It turns out that the essential spectrum of $H^\sharp_k(t)$ is independent of $t \in \R$, see {\em e.g.}~\cite{hempel2011variational, gontier2021edge}. This comes from the fact that the essential spectrum describes modes that escape to infinity, and that, far from the boundary, these modes only feel the bulk operator. In addition, for $t \in L \Z$, we recover the bulk spectrum, which is purely essential. Actually, we have
\begin{equation} \label{eq:sigma_ess}
   \forall t \in \R, \quad \sigma_{\rm ess} \left( H^\sharp_k(t) \right) = \sigma \left( H^\sharp_k(t = 0)  \right) = \bigcup_{n =1}^\infty \bigcup_{k_1 \in \left[\frac{-\pi}{L}, \frac{\pi}{L} \right]}  \{ \varepsilon_{n, \bk = (k_1, k)}\}.
\end{equation}

As $t$ varies, some additional eigenvalues may appear in the $t$-independent essential gaps, and we can define a {\em spectral flow} inside these gaps~\cite{atiyah1976spectral}. If $A(t)$ is a continuous $T$-periodic family of operators, and if $g$ is an open interval in an essential gap of all $A(t)$, we denote by
\[
    {\rm Sf} \left(A(\cdot), g, [0, T] \right)
\]
the spectral flow of $A(t)$ in the gap $g$, which counts the net number of eigenvalues going downwards in the gap $g$, when $t$ increases from $0$ to $T$. The following result is a reformulation of~\cite[Theorem 4.3]{hempel2011variational}.
\begin{theorem}[from~\cite{hempel2011variational}] \label{th:HK}
    For all $k \in\BZy$ and for all spectral gaps $g$ of $H^\sharp_k(t = 0)$, we have
    \[
        {\rm Sf} \left( H^\sharp_k(\cdot), g, [0, L] \right) = \cN(g).
   \]
\end{theorem}
Formally, when $t$ increases from $0$ to $L$, a new cell has appeared at the cut. Consider the state $\gamma_E(t) := \1(H_k^\sharp(t) < E)$ which describes a state with $\cN(E)$ particles per unit cell. The state $\gamma_E(t = L)$ must have $\cN(E)$ more particles than $\gamma_E(t = 0)$ in order to fill this new cell. These particles have been pumped from bands with higher energy, hence the presence of the spectral flow. While this reasoning is not accurate (we compare two infinities of particles), it describes the physics correctly. We refer to~\cite[Theorem 4.3]{hempel2011variational} and~\cite[Theorem 2.4]{hempel2012dislocation} for the full proof. In these works, the authors assumed $V$ to be Lipschitz in order to ensure that the branches of eigenvalues are continuous (actually Lipschitz). One can relax this assumption. We prove in Appendix~\ref{append:A} that the branches of eigenvalues are continuous whenever $V$ is bounded.

\subsection{Dirichlet Hamiltonian}

We now present similar results for the Dirichlet Hamiltonian $H^D$, which is acting on the half space $L^2(\R_2^+)$, where we set $\R^2_+ := \R_+ \times \R$. This operator is defined as
\[
    H^{D}(t) := - \Delta + V_t, \quad \text{acting on $L^2(\R^2_+)$, with domain $H^2(\R_+^2) \cap H^1_0(\R_+^2)$},
\]
that is with Dirichlet boundary conditions at the cut $\{x_1 = 0\}$. This operator still commutes with $L$-translations in the $x_2$-direction, and we may perform a partial Bloch theorem to write
\[
    H^{D}(t) = \int^\oplus_{\BZy} H^D_k(t) \rd k,
\]
where $H^D_k(t)$ acts on $L^2(\R_+ \times [0, L])$, with Dirichlet boundary conditions at the cut $\{ x_1 = 0\}$, and with $k$-quasi periodic boundary conditions in the $x_2$-direction. Explicitly, a core domain of $H_k^D(t)$ is given by
\begin{align*}
\big\{ & f(x_1, x_2) \in C^\infty( \R_+ \times [0, L] ),  \qquad \exists X > 0, \ \forall x_1 > X, \quad f(x_1, \cdot) = 0,  \\
   &  \quad f(0, \cdot) = 0, 
  \quad \forall \alpha \in \N_0, \forall x_1 \in \R_+ , \ \left( \partial_{2}^{\alpha} f \right)(x_1, L) = \re^{ \ri k L} \left( \partial_2^\alpha f \right)(x_1, 0) \big\}.
\end{align*}

\begin{lemma} \label{lem:ess_spectrum}
    For all $k \in \BZy$, and all $t \in \R$, we have
    \begin{equation} \label{eq:equality_ess_spectrum} 
        \sigma_{\rm ess} \left( H^D_k(t) \right) = \sigma_{\rm ess} \left( H^\sharp_k(t) \right).
    \end{equation}
    In particular, it is independent of $t \in \R$ by~\eqref{eq:sigma_ess}. In addition, for all spectral gaps $g$ of $H^\sharp_k(t = 0)$, we have
    \[
    {\rm Sf} \left( H^D_k(\cdot), g, [0, L] \right) = \cN(g).
    \]
\end{lemma}
\begin{proof}
    Introduce the operator $H_k^{\sharp, D}(t) = - \Delta + W_t$, which is similar to the dislocated operator $H_k^\sharp(t)$, but with Dirichlet boundary condition at the cut $\{ x_1 = 0\}$, that is with core domain (the function $f$ is now defined on the whole tube $\R \times [0, L]$)
    \begin{align*}
        \big\{ & f(x_1, x_2) \in C^\infty( \R \times [0, L] ), \qquad \exists X > 0, \ \forall | x_1 | > X, \quad f(x_1, \cdot) = 0, \\
        & \qquad  f(0, \cdot)= 0, 
        \quad \forall \alpha \in \N_0, \forall x_1 \in \R , \ \left( \partial_{2}^{\alpha} f \right)(x_1, L) = \re^{ \ri k L} \left( \partial_2^\alpha f \right) (x_1, 0) \big\}.
    \end{align*}
    Let $\mu$ be a large negative number so that $\mu < \inf \sigma(H_k^{\sharp, D})$ and $\mu < \inf \sigma (H_k^\sharp)$, {\em e.g.} $ \mu := - \| V \|_\infty - 1$. It is a classical result that  (see {\em e.g.}~\cite[Theorem 1.1]{carron2002determinant}, ~\cite[Theorem XI.79]{reed1980methods} or the discussion before~\cite[Theorem 2.4]{hempel2012dislocation})
    \begin{equation} \label{eq:diff_res_compact}
        \left(H_k^{\sharp, D}(t) - \mu \right)^{-1} - \left(H_k^{\sharp}(t) - \mu \right)^{-1} \quad \text{is compact}.
    \end{equation}
This already proves that these operators have the same essential spectrum. In addition, since the spectral flow is robust with respect to perturbation with compact operators (see {\em e.g.}\footnote{This Proposition states that Sf($\cdot$) is a homotopy invariance for the so-called gap topology. Let $(A_t)$ be a continuous periodic family of self-adjoint operators  sharing an essential gap $g$, and let $(K_t)$ be a continuous periodic family of compact operators. For $s \in [0, 1]$, consider the path
    \[
    C_s(t) := \begin{cases}
        A_0 + 3 t s K_0 & \quad 0 \le t \le 1/3 \\
        A_{3(t - 1/3)} + s K_{3(t - 1/3)} & \quad 1/3 \le t \le 2/3 \\
        A_1 + 3(1 -t) s K_1 & \quad 2/3 \le t \le 1.
    \end{cases}
    \]
    Since $K_t$ is compact, this path is continuous for the gap topology. By periodicity, $A_0 = A_1$ and $K_0 = K_1$, so the contribution of $C_s$ to the spectral flow for $t \in [0, 1/3]$ cancels with the one for $t \in [2/3, 1]$. Also, by continuity, $s \mapsto {\rm Sf}(C_s(\cdot), g)$ is continuous and integer-valued, hence constant. We obtain
    \[
        {\rm Sf}(A_t, g) = {\rm Sf}(C_s(t), g) = {\rm Sf}(t \mapsto A_t + s K_t, g) = {\rm Sf}(B_t, g), \quad \text{with} \quad B_t = A_t + K_t.
    \]
}~\cite[Proposition 3]{phillips1996self}), we have
    \[
      {\rm Sf} \left( \left( H^{\sharp, D}_k(\cdot) - \mu \right)^{-1}, (g - \mu)^{-1}, [0, L] \right) = {\rm Sf} \left( \left( H^{\sharp}_k(\cdot) - \mu \right)^{-1}, (g - \mu)^{-1}, [0, L] \right).
    \]
    An eigenvalue of $(A(\cdot) - \mu)^{-1}$ crosses the energy $(E - \mu)^{-1}$ upwards iff an eigenvalue of $A(\cdot)$ crosses the energy $E$ downwards. We deduce that
    \[
       {\rm Sf} \left( H^{\sharp, D}_k(\cdot), g , [0,L] \right) = {\rm Sf} \left( H^\sharp_k(\cdot), g , [0, L] \right) =  \cN(g) .
    \]
    We used Theorem~\ref{th:HK} in the last equality. For the operator $H^{\sharp, D}_k(t)$, the left and right channels are decoupled by the Dirichlet boundary conditions. Since the left channel is independent of $t$, it does not contribute to the spectral flow. On the right channel, we recover the Dirichlet Hamiltonian $H_k^D(t)$ on the semi-tube, and the result follows.
\end{proof}


\section{Application to half twisted Hamiltonians}

We now apply the previous theory in the case of the twisted Hamiltonian $H[\theta]$. 

\subsection{Spectrum for rational angles}
We first assume that $\tan(\theta)$ is rational, of the form $\tan \theta = \frac{p}{q}$ with $p$ and $q$ relatively prime, and we set $L := (p^2 + q^2)^{1/2}$. The matrix $R_\theta$ can be written as
\[
    R_\theta = \frac{1}{\sqrt{p^2 + q^2}} \begin{pmatrix}
        q & -p \\ p & q
    \end{pmatrix},
\quad \text{and} \quad
R_\theta^{-1} = \frac{1}{\sqrt{p^2 + q^2}} \begin{pmatrix}
    q & p \\ -p & q
\end{pmatrix}.
\]
In particular, since $V$ is $\Z^2$-periodic, $V_\theta(x)= V(R_\theta^{-1} x )$ is $L \Z^2$-periodic.

Let $E$ be in the resolvent set of $H$, and let $\cN(E)$ be the number of Bloch bands below $E$, when $H$ is seen as a $\Z^2$-periodic operator. The operator $\gamma_E := \1(H < E)$ represents a state with $\cN(E)$ particles per unit cell. Since $H[\theta]$ is a rotation version of $H$, the energy $E$ is also in the resolvent set of $H[\theta]$. Seeing $H[\theta]$ as an $L\Z^2$-periodic operator (with unit cell of area $L^2 = p^2 + q^2$), there are 
\[
     \cN_\theta(E) := L^2 \cN(E) = (p^2 + q^2) \cN(E)
\] 
Bloch bands below $E$ for this operator. Applying the results of the previous section, we obtain the following. We denote by $H^\sharp[\theta, t]$, $H^D[\theta, t]$ the $t$-dislocated and Dirichlet version of the operator $H[\theta]$ respectively.
\begin{lemma}
    Assume $\tan \theta = \frac{p}{q} \in \Q$, and set $L := \sqrt{p^2 + q^2}$. Then, for all essential gap $g \subset \R \setminus \sigma_{\rm bulk}$, and all $k \in \BZy = \left[ - \frac{\pi}{L}, \frac{\pi}{L}  \right]$, we have
    \[
        {\rm Sf} \left( H^{\sharp}_k[ \theta, \cdot ], g, [0, L]  \right) = {\rm Sf} \left( H^{D}_k[ \theta, \cdot ], g, [0, L]  \right) = L^2 \cN(E).
    \]
\end{lemma}

Actually, the Dirichlet operators $H^{D}_k[\theta, t]$ and $H^{D}_k \left[ \theta, t + \frac{1}{L}\right]$ have the same spectrum. One way to see this goes as follows. In the cell $[0, L)^2$, there are $L^2$ points from the grid $R_\theta \Z^2$. When $t$ swipes from $0$ to $L$, each one of these points crosses the line $\{0\} \times \R$, for a total of $L^2$ crossings. By periodicity, these crossings are regularly spaced, with spacing $1/L$. When one of them occurs, we recover the operator $H^{D}_k[\theta, 0]$, translated in the $x_2$-direction. So, we may formally write
\begin{equation} \label{eq:sf_Dirichlet}
   {\rm Sf} \left( H^{D}_k[ \theta, \cdot ], g, [0, L^{-1}]  \right) = \cN(E).
\end{equation}
We deduce the following side result.

\begin{lemma}
    Assume $V$ is $\nu$-Lipschitz. Then, for all $t \in \R$, for all $k \in \BZy$ and for all gaps $g = (a,b)$ of $\sigma_\bulk$, there are at least
    \[
       \left\lfloor \frac{(b - a)L}{\nu} \right\rfloor \cN(E) 
    \]
    eigenvalues of $H^D_k[\theta, t]$ in this gap, counting multiplicities.
\end{lemma}

So, when $L \to \infty$, the gap $g$ is filled with eigenvalues. This is already an indication that the gap $g$ is completely filled for $\tan \theta \notin \Q$. We prove this fact in the next section.

\begin{proof}
    If $\cN(E) \neq 0$, there is $t_0 \in [0, L^{-1})$ so that $H_k^D[\theta, t_0]$ has at least one eigenvalue $\lambda_0$ in $g$. Without loss of generality, we may assume $t_0 = 0$. We write 
    \[
        \cdots \le \lambda_{-1}(t) \le \lambda_0(t) \le \lambda_1(t) \le \cdots
    \]
    the continuous branches of eigenvalues of $H_k^D(t)$ counting multiplicities, and with $\lambda_0(0) = \lambda_0$. These functions are defined for all $t \in \R$, with the convention $\lambda_j(t)= a$ (resp. $b$) if the branch merges to the lower (resp. upper) essential band. There may be a finite or a countable number of such branches.
    
    Since $V$ is $\nu$-Lipschitz, we have $\| V_{t'} - V_{t} \|_\infty \le \nu | t' - t |$. In particular, for $\mu$ negative enough, the map $t \mapsto \left( \mu - H_k^D(t) \right)^{-1}$ is Lipschitz for the operator norm, and we deduce that the functions $\lambda_j(\cdot)$ are also $\nu$-Lipschitz (see also~\cite[Theorem 1.2]{hempel2015bound}). So we have $| \lambda_0(L^{-1}) - \lambda_0(0) | \le \nu/L$. By periodicity of the spectrum, $\lambda_0(L^{-1})$ also corresponds to an eigenvalue of $H_k^D[t = 0]$ (unless this branch already merged to the essential spectrum). Eqn.~\eqref{eq:sf_Dirichlet} states that $\lambda_0(L^{-1}) = \lambda_{-\cN(E)}(0)$. We deduce that there are at least $\cN(E)$ eigenvalues in any interval of size $\nu/L$. There are $\lfloor(b - a)L/\nu \rfloor$ disjoint intervals of this size in $g = (a,b)$, and the result follows.
\end{proof}

\subsection{The spectrum for irrational angles}
\label{sec:proof}

We finally consider the case where $\tan \theta \notin \Q$. We want to study the spectrum of $H^D[\theta, t]$. Since $\tan \theta \notin \Q$, this spectrum is independent of $t \in \R$ by ergodicity.

The main idea is to approximate $\theta$ by a sequence $\theta_n \to \theta$ with $\tan \theta_n = \frac{p_n}{q_n} \in \Q$, and to control the corresponding eigenstates in suitable spaces. We set $L_n := \sqrt{p_n^2 + q_n^2}$.

{\bf Step 1: Construction of a subsequence.}
Fix $E$ an energy in a gap $g$ of $\sigma_\bulk$, and above the first Bloch band ($\cN(E) \ge 1$). By Eqn.~\eqref{eq:sf_Dirichlet} at $k = 0$, there is $t_n \in [0, L_n^{-1}]$ and a wave-function $\Psi_n(x_1, x_2)$ in the domain of $H_{k = 0}^D$ with
\[
    H_{k = 0}^D[\theta_n, t_n] \Psi_n = E \Psi_n.
\]
We may assume that $\Psi_n$ is real-valued. \review{The main idea of the proof is to normalize $\Psi_n$ for the $L^\infty$ norm, that is with $\| \Psi_n \|_{L^\infty} = 1$. This idea was used for instance in~\cite[Theorem 3.6]{avron1983almost} in the context of almost periodic one-dimensional operators. 

We have
\[
    \| (- \Delta) \Psi_n \|_\infty = \| (E - V_{\theta_n, t_n} ) \Psi_n \|_\infty \le \| V \|_\infty + | E |,
\]
so $\| (- \Delta) \Psi_n \|_\infty$ is also uniformly bounded in $n$. In addition, by elliptic regularity, we have $\Psi_n \in W^{2,p}_{\rm loc}$ for all $1 < p < \infty$, and, if $Q_{ij}$ is the square $Q_{ij} := (i, i+2) \times (j,j+2)$, we have $\| \Psi_n \|_{W^{2,p}(Q_{ij})} \le C'$ for some constant $C'$ independent of $(i,j)$ and of $n$. Together with the Morrey and Sobolev embeddings, we deduce that $\Psi_n$ is in $\cC^{1, \alpha}$ for all $ 0 \le \alpha < 1$, and that $\| \nabla \Psi \|_{L^\infty} \le C''$ for a constant $C''$ independent of $n$.}

We now extend $\Psi_n$ by periodicity in the $x_2$-direction to obtain a function on the whole half plane $\R^2_+$, still denoted by $\Psi_n$, and for which
\[
    \begin{cases}
        \| \Psi_n \|_{L^\infty(\R^2_+)} = 1, \\
        \| (- \Delta) \Psi_n \|_{L^\infty(\R^2_+)} \le C, \\
        \| \nabla \Psi_n \|_{L^\infty(\R^2_+)} \le C'', \\
        ( - \Delta + V_{\theta_n, t_n} - E) \Psi_n = 0 \quad \text{in the distributional sense}.
    \end{cases}
\]

We would like to extract a subsequence which converges to $\Psi_*$ weakly-* in $L^\infty(\R^2_+)$. However, should we take limits directly, one could end up with $\Psi_* = 0$ at the limit. We first need to control that the mass of $\Psi_n$ does not escape. 

\medskip

{\bf Step 2: Controlling the mass "vertically".}
Let $\bx_n \in \R^2_+$ be a point for which $| \Psi_n |(\bx_n) \ge \frac12$. Up to multiplying $\Psi_n$ by a global sign (recall that $\Psi_n$ is real-valued), we may assume $\Psi_n (\bx_n) \ge \frac12$. We now translate the whole system vertically, in order to put the point $\bx_n$ on the horizontal semi-line $\{ x_2 =  0 \}$. This will guarantee that some mass of $\Psi_n$ stays around the this semi-line. More specifically, we have
\[
    \Big( - \Delta + V( R_{\theta_n}^{-1} \left( \bx - t_n \be_1 - x_{2,n} \be_2 \right))  - E \Big) \Psi_n (\cdot - x_{2,n} \be_2)= 0.
\]
We introduce the vector $\bt_n \in R_{\theta_n} [0, 1]^2$ so that
\[
    \bt_n := t_n \be_1 + x_{2,n} \be_2 \quad {\rm mod} \ R_{\theta_n} \Z^2.
\]
By periodicity of $V$, and by setting $V_{\theta_n, \bt_n}(\bx) := V_{\theta_n}(\bx - \bt_n)$, we have
\[
    ( - \Delta + V_{\theta_n, \bt_n }  - E ) \widetilde{\Psi_n} = 0,
\]
where $ \widetilde{\Psi_n} (\bx) := \Psi_n(\bx- x_{2,n} \be_2)$ is a $x_2$-translated version of $\Psi_n$. Setting $\widetilde{\bx_n} = \bx_n - x_{2, n} \be_2 = (x_{1,n}, 0)$, the point $\widetilde{\bx_n}$ belongs to the semi-line $\R_+ \times \{ 0 \}$, and we have $\widetilde{\Psi_n}(\widetilde{\bx_n}) \ge \frac12$ for all $n$. In what follows, we drop the tilde notation, and write $\Psi_n$ and $\bx_n$ for $\widetilde{\Psi_n}$ and $\widetilde{\bx_n}$.

\medskip

{\bf Step 3. Controlling the mass "horizontally".} Let $\chi_1(x_1, x_2) = \chi_1(x_1)$ be a smooth switch function in the $x_1$ direction, such that $\chi_1(x_1) = 0$ for $x_1 < \frac12$ and $\chi_1(x_1) = 1$ for $x_1 > 1$. We introduce the function
\[
    f_{n} :=  (- \Delta + V_{\theta_n, \bt_n} - E) \left( \chi_1  \Psi_n \right) = - 2 \chi_1' (\partial_{1} \Psi_n) -  \Psi_n \chi_1''.
\]
This function has support in the vertical band $(0, 1) \times \R$, and is bounded by
\[
    \| f_n \|_\infty \le 2 \| \chi_1' \|_\infty \| \partial_1 \Psi_n \|_\infty + \| \chi_1 '' \|_\infty \| \Psi_n \|_\infty  \le C ,
\]
for some constant $C > 0$ independent of $n$. 

Let $G_{n}(\bx, \by)$ be the kernel of the bulk resolvent $(H[\theta_n, \bt_n] - E)^{-1}$. Since $E \in \sigma_\bulk$, the Combes-Thomas argument (see~\cite{combes1973asymptotic} and~\cite[Theorem B.7]{simon1982schrodinger}) states that there is $C \ge 0$ and $\alpha, \alpha' > 0$, independent of $n$, so that
\[
    | G_n (\bx, \by) | \le C \re^{- \alpha' | \bx - \by |} \le C \re^{ - \alpha | x_1 - y_1 | }  \re^{ - \alpha | x_2 - y_2 | }.
\]
Let $x_1 > 1$ and set $\bx := (x_1, 0)$. We have
\begin{align*}
    \Psi_n(\bx)  & = \chi_1(\bx) \Psi_n(\bx) = \big[ (H[\theta_n, \bt_n] - E)^{-1} (- \Delta + V_{\theta_n, \bt_n} - E) \left(\chi \Psi_n \right) \big](\bx) \\
    & = \int_{\R^2} G_n(\bx, \by) f_n(\by) \rd \by =  \int_{[0, 1] \times \R} G_n(\bx, \by) f_n(\by) \rd \by.
\end{align*}
This gives
\[
    \left|  \Psi_n \right|(\bx) \le \int_{[0, 1] \times \R} | G_n | (\bx, \by) \| f_n \|_\infty \rd \by \le C \int_{[0, 1]} \re^{- \alpha | x_1 - y_1|} \rd y_1 \le C \re^{ - \alpha | x_1 - 1 | }.
\]
In other words, $\Psi_n$ is exponentially localized near the cut. We deduce that there is $X > 1$ so that, for all $x_1 > X$, we have $| \Psi_n (x_1, 0) | < 1/2$. In particular, the point $\bx_n$ belongs to the (compact) segment $[0, X] \times \{ 0 \}$. In addition, since $(\nabla \Psi_n)$ is bounded, there is $\delta > 0$ independent of $n$ so that $\Psi_n (\bx) > \frac14$ for all $\bx$ in the ball $\cB_\delta(\bx_n) :=  \{ \bx \in \R^2_+, \  |\bx - \bx_n | < \delta\}$.

\medskip

{\bf Step 4. Extracting a subsequence.} At this point, the sequence $(\bt_n)$ is bounded in $\R^2$, the sequence $(\bx_n)$ belongs to the compact set $[0, X] \times \{ 0 \}$, and $(\Psi_n)$ and $(-\Delta \Psi_n )$ are bounded in $L^\infty(\R^2_+)$. Up to a subsequence, still denoted by $n$, we may assume that
\[
    \bt_n \xrightarrow[n \to \infty]{} \bt_*, \quad \bx_n \xrightarrow[n \to \infty]{} \bx_*, \quad 
    \begin{cases}
        \Psi_n \xrightarrow[n \to \infty]{} \Psi_* & \ \text{weakly-* in $L^\infty(\R^2_+)$}, \\
        (-\Delta) \Psi_n \xrightarrow[n \to \infty]{} T & \ \text{weakly-* in $L^\infty(\R^2_+)$}.
    \end{cases}
\]
Since $\Psi_n \to \Psi_*$ in the distributional sense, we have $T = (- \Delta) \Psi_*$.
In addition, we have $\Psi_*(\bx) > \frac14$ on $\cB_\delta(\bx_*)$ so $\Psi_*$ is non-null. It remains to prove that $(-\Delta + V_{\theta, \bt_*} - E) \Psi_* = 0$ in the distributional sense. For $\phi \in C^\infty_0(\R^2_+)$ a test function, we have, 
\[
    0 = \bra \phi, (- \Delta + V_{\theta_n, \bt_n} - E) \Psi_n \ket = \bra (- \Delta - E) \phi, \Psi_n \ket + \int_{\R^2_+} \phi V_{\theta_n, \bt_n} \Psi_n.
\]
The first term goes to $\bra (-\Delta - E) \phi, \Psi_*  \ket$ by the weak-* convergence of $\Psi_n$ to $\Psi_*$. For the second term, we get, after changing variables in order to put the rotation on the test functions,
\begin{align*}
   &  \int \left| \phi \left( V_{\theta_n, \bt_n} \Psi_n - V_{\theta, \bt_*} \Psi_* \right) \right|  = 
    \int \left| V(\bx) \left( [ \phi \Psi_n ] (R_{\theta_n} \bx + \bt_n)  -  [ \phi \Psi_* ] (R_{\theta} \bx + \bt_*) \right)(\bx) \right| \\
    & \qquad  \le \| V \|_\infty \int \left| [ \phi \Psi_n ] (R_{\theta_n} \bx + \bt_n)  -  [ \phi \Psi_* ] (R_{\theta} \bx + \bt_*) \right| \rd \bx \\
    & \qquad \le \| V \|_\infty \left( \int  \left| \phi (\Psi_n - \Psi_*) \right| + \int \left| [\phi \Psi_*](R_{\theta_n} \bx + \bt_n ) -  [\phi \Psi_*](R_{\theta} \bx + \bt )  \right| \right).
\end{align*}
Note that all integrals are set on compact supports. Since rotations and translations are continuous operators in all $L^p$ spaces $1 \le p < \infty$, together with the fact that $\Psi_n \to \Psi$ strongly in $L^p_{\rm loc}$ for all $1 \le p < \infty$ by Rellich embedding, we conclude that this term vanishes as well as $n \to \infty$. This proves that
\[
    \bra \phi, ( - \Delta + V_{\theta, \bt^*} - E) \Psi_* \ket = 0, \quad \text{for all $\phi \in C^\infty_0(\R^2_+)$,}
\]
hence $\Psi_*$ is a distributional solution to $(-\Delta + V_{\theta, \bt_*} - E) \Psi_* = 0$, which belongs to $L^\infty(\R^2_+)$. We conclude that $E \in \sigma \left( H[\theta, \bt_*] \right) = \sigma \left( H[\theta] \right)$, as wanted.

\subsection{Domain wall Hamiltonians} \label{ssec:domain_wall}

Our proof also allows to treat the case of half materials set in the whole space $\R^2$. Let $\chi$ be a bounded switch function with $\chi(\bx) = 0$ for $x_1 < X$ and $\chi(\bx) = 1$ for $x_1 > X$, where $X \ge 0$ is some large number. We study the {\em domain wall} edge operator
\begin{equation} \label{eq:domain_wall}
    H^\chi[\theta] := - \Delta + \chi(\bx) V_{\theta}(\bx ).
\end{equation}
The potential $\chi V_\theta$ vanishes on the left, and equals $V_\theta$ on the right, so $H^\chi[\theta]$ models a semi-material embedded in the full $\R^2$ space, and cut with an angle $\theta$. Again, if $\tan \theta = \frac{p}{q} \in \Q$ is rational, this operator is periodic in the $x_2$-direction (with period $(p^2 + q^2)^{1/2}$), and therefore has the bulk-gap spectrum. In the incommensurable case $\tan \theta \notin \Q$, the counterpart of Theorem~\ref{th:main} reads as follows.
\begin{theorem} \label{th:DW}
    If $\tan \theta \notin \Q$, then there is $\Sigma \in \R$ such that $\sigma(H^\chi[\theta]) = [\Sigma, \infty)$.
\end{theorem}

The main difference with the Dirichlet case is that the left side is now the free Laplacian $-\Delta$, whose essential spectrum is $[0, \infty)$. In particular, the counterpart of Eqn.~\eqref{eq:equality_ess_spectrum} is
\[
    \sigma_{\rm ess} \left( H^\chi_k(t) \right) = \sigma_{\rm ess} \left( H^\sharp_k(t = 0) \right) \cup [0, \infty).
\]
So the bulk gaps with positive energy are filled with the essential spectrum of the free Laplacien $(-\Delta)$, and the lower energy bulk gaps are filled with edge spectrum if $\tan \theta \notin \Q$. We leave the details of the proof of Theorem~\ref{th:DW} to the reader, as it is similar to the Dirichlet case.

\appendix

\section{Continuity of the branches of eigenvalues}
\label{append:A}

In this section, we prove that the eigenvalues of $H^\sharp_k(t)$ are continuous in $t$, under the sole assumption that $V$ is bounded. This extends~\cite[Theorem 1.2]{hempel2015bound}. Recall that 
\[
    H^\sharp (t) = - \Delta + W_t \quad \text{acting on $L^2(\R^2)$ with domain $H^2(\R^2)$},
\]
where $W_t$ is the dislocated potential in Eqn.~\eqref{eq:def:Wt}, and that $H^\sharp_k(t)$ are the (partial) Bloch fibers of $H^\sharp(t)$, acting on $L^2(\R \times (0,L))$.  In what follows, we consider an essential gap $g = (a, b)$ for $H^\sharp_k(t = 0)$ (hence for all $H^\sharp_k(t)$). 

Let us first prove that $t \mapsto H_k^\sharp(t)$ is continuous for the strong resolvent topology~\cite[Chapter VIII.7]{reed2012methods}. All operators $H^\sharp_k(t)$ share the same core domain
\begin{align*}
    \big\{  & f(x_1, x_2) \in C^\infty( \R \times [0, L] ), \qquad  
    \exists X > 0, \quad \forall | x_1 | > X, \quad f(x_1, \cdot) = 0, \\
    & \qquad \forall \alpha \in \N_0,  \ \left( \partial_{2}^{\alpha} f \right)(\cdot, L) = \re^{ \ri k L} \left( \partial_2^\alpha f \right)(\cdot, 0) \big\}.
\end{align*}
Let $u$ be in this domain, and set $\tilde{u}(\bx) := u(\bx) \1(x_1 > 0)$. We have
\begin{align*}
    \left\| \left( H^\sharp_k(t+h) - H^\sharp_k(t)  \right) u \right\|_{L^2}^2 & = \int_{\R \times (0, L)}  (V_{{t+h}} - V_t)^2  | \tilde{u} |^2.
\end{align*}
Since $\tilde{u}$ is compactly supported, there is $X > 0$ large enough so that $[-X, X] \times [0, L]$ contains the support of $\tilde{u}(\cdot + t \be_1)$ and $\tilde{u}(\cdot + (t + h) \be_1)$ for all $ | h | \le 1$. Setting $\widetilde{V}(\bx) := V(\bx) \1( | x_1 | \le X)$, and $M := \max | \tilde{u} |$, we obtain, for $| h | \le 1$,
\[
    \left\| \left( H^\sharp_k(t+h) - H^\sharp_k(t)  \right) u \right\|_{L^2}^2  = \int_{\R \times (0, L)}  (\widetilde{V}_{{t+h}}  - \widetilde{V}_t)^2  | \tilde{u} |^2 \le M^2 \left\| \widetilde{V}_{{t+h}}  - \widetilde{V}_t \right\|_{L^2}^2.
\]
The function $\widetilde{V}$ is bounded and compactly supported, hence belongs to $L^2$. Since translations are continuous in $L^2$, this term goes to $0$ as $h \to 0$. We deduce that $H^\sharp_k(t+h) u$ converges to $H^\sharp_k(t)u$ as $h \to 0$. Together with~\cite[Theorem VIII.25]{reed2012methods}, this proves that $t \mapsto H^\sharp(t)$ is strongly resolvent continuous. In particular, {\em gaps cannot suddenly expand} (see~\cite[Chapter VIII \S1.2]{kato2013perturbation}).

Next, we prove that {\em gaps cannot suddenly shrink}, in the sense that, for all $t \in \R$ and all energies $E \in (a, b) \setminus \sigma(H^\sharp_k(t))$, we have
\begin{equation*} 
  \exists \eta, \eps > 0, \quad
    \forall t' \in (t-\eta, t + \eta), \quad \sigma (H^\sharp_k(t')) \cap (E-\eps, E + \eps) = \emptyset.
\end{equation*}
Assume otherwise, and let $t_n \to t$ and $\lambda_n \to E$ be sequences so that $\lambda_n \in \sigma (H^\sharp_k(t_n))$. Recall that $\sigma(H^\sharp_k(t)) \cap (a,b)$ only consists of eigenvalues. So there are $u_n \in H^2(\R \times (0, L))$ with
\[
    (- \Delta + W_{t_n}) u_n = \lambda_n u_n.
\]
We normalize $(u_n)$ in the way $\| u_n \|_{L^\infty} = 1$. The family $(u_n)$ is bounded in $L^\infty$, hence converges weakly-$*$ to some $u$ in $L^\infty(\R \times (0, L))$ up to a subsequence. We can now repeat the arguments of Section~\ref{sec:proof} to deduce that $u$ does not vanish almost everywhere, and satisfies
\[
    (-\Delta + W_{t}) u = E u.
\]
We deduce that $E \in \sigma(H^\sharp_k(t))$, a contradiction. 

Finally, let $\lambda \in \sigma \left( H^\sharp_k(t) \right)$ be an isolated eigenvalue of multiplicity $m$, and let $\sC$ be a small positively oriented loop in the complex plane enclosing $\lambda$ and no other eigenvalue of $H^\sharp_k(t)$. By the stability of the gaps, there is $\eta > 0$ so that the spectrum of $H^\sharp_k(t')$ does not touch $\sC$ for all $t' \in (t-\eta, t + \eta$). We set
\[
 \forall t' \in (t-\eta, t + \eta), \quad    P_{t'} := \frac{1}{2 \ri \pi} \oint_{\sC} \dfrac{\rd z}{z- H^\sharp_k(t') }.
\]
By the Cauchy residual formula, this defines a family of projectors. Reasoning as before (with a family of orthogonal functions), we can prove that there is $\eta' > 0$ so that, for all $t' \in (t - \eta', t + \eta')$, we have $\dim \ P_{t'} \le m$. We are now in the setting of~\cite[Chapter VIII \S1.4]{kato2013perturbation} and we conclude that the branches of eigenvalues are continuous.



\gdef\og{``}\gdef\fg{''}
\providecommand\cdrnumero{no.~}
\providecommand{\cdredsname}{eds.}
\providecommand{\cdredname}{ed.}
\providecommand{\cdrchapname}{chap.}
\providecommand{\cdrmastersthesisname}{Memoir}
\providecommand{\cdrphdthesisname}{PhD Thesis}


\end{document}